# Comment on Zarechnaya et al. Pressure-induced isostructural phase transformation in γ-B$_{28}$. *Phys. Rev. B* 82, 184111 (2010)

by Yann Le Godec[1]*, Artem R. Oganov[2], Oleksandr O. Kurakevych[1], Vladimir L. Solozhenko[3]

[1] *Institut de Minéralogie et de Physique des Milieux Condensés, CNRS, Université P&M Curie, 75015 Paris, France*
[2] *Department of Geosciences and New York Center for Computational Science, Stony Brook University, Stony Brook, New York 11794, USA*
[3] *Laboratoire des Sciences des Procédés et des Matériaux, CNRS, Université Paris Nord, Villetaneuse, 93430, France*

γ-B$_{28}$, an allotrope of boron whose crystal structure, stability field, and physical properties have been established in the last several years, has attracted tremendous attention.[1-8] Of particular interest are the mechanical properties, such as the superhardness [2,4] and equation of state (EOS) [4-8] of this phase. The EOS of γ-B$_{28}$ was measured experimentally at 0-65 GPa [4], 0-30 GPa [5], 0-60 GPa [6] and 0-40 GPa [7] using diamond anvil cells and X-ray diffraction with synchrotron radiation. Theoretical calculations at the DFT-GGA level ($B_0$=241 GPa [7] and $B_0'$=2.34 [7] at 300 K) agree well with some experiments ($B_0$=237 GPa [4,7] and $B_0'$=2.7 [4] or 2.5 [7]) but disagree with other experiments ($B_0$=227 GPa [5] and $B_0'$=2.2 [5]). Zarechnaya *et al.*[6] claimed an isostructural transformation in γ-B$_{28}$ at about 40 GPa; below which the phase is more compressible ($B_0$=227 GPa [6]) and above which less compressible ($B_0$=281 GPa [6]) than in previous experiments [4] or theory [5-8]. Here we wish to point out some interesting questions related to these claims. Summarizing briefly our points, the suggestion of an isostructural transformation is inconsistent with *ab initio* calculations [2,7,8] and experiment [4], we see no physical mechanism that could be responsible for such a transformation, and the evidence for it reported in Ref. 6 is self-contradictory.

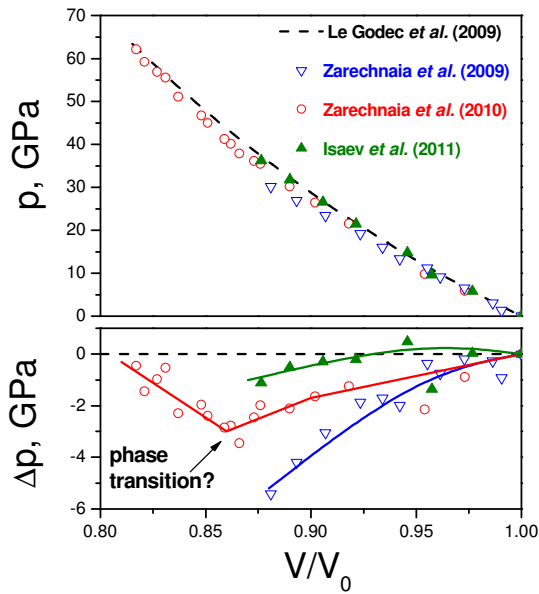

Fig. 1. Experimental equations of state of γ-B$_{28}$ (colored online). (*Top*) Experimental data of different authors[5-7] (symbols) in comparison with the data by Le Godec *et al.*[4] (dashed line). (*Bottom*) Difference between the experimental equations of state[5-7] (the symbols represent the same data as on the top) and that of Le Godec *et al.*[4] (dashed line). The solid lines are guides to the eye.

Isostructural phase transitions/transformations[9] (a subset of more general isosymmetric transformations, occurring with no change of space group) are well known in many systems and, while preserving symmetry and structure type, involve discontinuous changes in the density, electronic structure, or even coordination numbers (if allowed by space group), all of which imply discontinuities in all vibrational frequencies – while experiments of Zarechnaya *et al.*[6] observed weak discontinuities only of some Raman modes (but not of the density, electronic structure or coordination numbers), and continuous behavior of the rest. Furthermore, only continuous behavior of Raman frequencies was reported in their *ab initio* calculations at zero temperature.

As has been shown[10] using Landau theory, in agreement with experimental evidence, isostructural phase transformations must be first-order below the critical temperature and fully continuous above it. This involves a soft symmetry-preserving A$_g$ mode at the observed transformation pressure. No such soft mode and no isostructural phase transformation have been seen by any calculations done on γ-B$_{28}$ to date. Recent calculations of Isaev *et al.*[7] explicitly show the absence of any mode softening (instead, showing mode hardening) at 40 GPa both at 0 K and 300 K. Ref. 6 argued that the isostructural transformation is related to a kink in the pressure dependence of the LO-TO splitting parameter ξ, which according to Ref. 6 characterizes the degree of polarity (i.e. ionicity) of bonding. In reality, the relationship of bond polarity and ξ in crystals is not direct – ξ originates from dynamical (rather than static) charges. If bond ionicity is negligible (as claimed by Zarechnaya *et al.* earlier[5]) and only dynamical charges are significant, the ξ(*P*) dependence can affect only the phonon part of the equation of state, as can be easily shown, shifting it by not more than $\Delta P_{max} = \frac{B_T E_{vib}}{6NV} \frac{d\ln\xi}{dP}$, where $B_T$, $E_{vib}$, $V$ and $N$ are the isothermal bulk modulus, vibrational energy, volume, and number of atoms in the unit cell, respectively. These values can be extracted from Refs. 1,3-7, yielding negligibly small $\Delta P_{max} <$ 0.02 GPa, two orders of magnitude smaller than a typical uncertainty of pressure measurements. It remains to see why experiments of Zarechnaya *et al.*[6] observed an isostructural phase transformation, while other experiments and *ab initio* calculations provide no evidence for it. The discrepancies between experiment and theory are always stimulating, and we would like to see further works clarifying the nature or the absence of the isostructural phase transformation observed[6] in γ-B$_{28}$. The view that emerges from the majority of evidence, including the most recent theoretical-experimental work of Isaev *et al.*[7], is that DFT-GGA is capable of very accurately describing the EOS of γ-B$_{28}$ and other boron allotropes, and that there is no isostructural phase transformation in γ-B$_{28}$ within its stability field. The isostructural phase transformation is likely to be an artifact of experiments of Zarechnaya *et al.*[6].